\documentclass[aps,prl,reprint,groupedaddress,citeautoscript,footinbib]{revtex4-1}

\usepackage{graphicx}
\usepackage{amsmath}
\usepackage{amssymb}
\usepackage{amsfonts}
\usepackage{mathrsfs}
\usepackage{bm}
\usepackage{array}
\newcolumntype {s}[1]{@{\hspace{#1}}} 
\usepackage{color}

\graphicspath{{.}{./EPS/}}

\newcommand* {\ee}{\ensuremath{\mathrm{e}}}

\newcommand*{\vek}[1]{{\ensuremath{\bm{\mathrm{#1}}}}}

\begin{document}

\title{Carrier-Density-Controlled Anisotropic Spin Susceptibility of Two-Dimensional Hole
Systems}

\author{T. Kernreiter}
\affiliation{School of Chemical and Physical Sciences and MacDiarmid Institute for Advanced
Materials and Nanotechnology, Victoria University of Wellington, PO Box 600, Wellington
6140, New Zealand}

\author{M. Governale}
\affiliation{School of Chemical and Physical Sciences and MacDiarmid Institute for Advanced
Materials and Nanotechnology, Victoria University of Wellington, PO Box 600, Wellington
6140, New Zealand}

\author{U. Z\"ulicke}
\affiliation{School of Chemical and Physical Sciences and MacDiarmid Institute for Advanced
Materials and Nanotechnology, Victoria University of Wellington, PO Box 600, Wellington
6140, New Zealand}

\begin{abstract}
We have studied quantum-well-confined holes based on the Luttinger-model description
for the valence band of typical semiconductor materials. Even when only the lowest
quasi-two-dimensional (quasi-2D) subband is populated, the static spin susceptibility turns
out to be very different from the universal isotropic Lindhard-function lineshape obtained for
2D conduction-electron systems. The strongly anisotropic and peculiarly density-dependent
spin-related response of 2D holes at long wavelengths should make it possible to switch
between easy-axis and easy-plane magnetization in dilute magnetic quantum wells. An
effective \textit{g} factor for 2D hole systems is proposed.
\end{abstract}

\date{\today}


\maketitle

\textit{Introduction} --
In semiconductors, electric current can be carried by conduction-band electrons
or valence-band holes. Besides being oppositely charged, these two types of
quasiparticles exhibit strikingly different magnetic properties. Band electrons are
spin-1/2 particles like electrons in vacuum. In contrast, holes have a spin angular
momentum of 3/2. Size quantization strongly affects the holes'
spin-3/2 degree of freedom~\cite{rolandbook}. This is believed to be the origin
of an unusual paramagnetic response observed for quantum-well-confined
holes~\cite{win05a,bou06,chi11}, and the same phenomenon is expected to stabilize
out-of-plane easy-axis magnetism in a dilute-magnetic-semiconductor
(DMS)~\cite{ohno:sci:98,jung:rmp:06,dietl:nmat:10} two-dimensional (2D) hole
system~\cite{haury:prl:97,dietl:prb:97}. Controlling the confinement of holes thus
enables appealing routes toward realizing magnetic semiconducting devices based on
strain-induced anisotropies~\cite{papp:nphys:07} or wave-function engineering in
heterostructures~\cite{naz:prl:05,wurst:nphys:10}. Here we present a detailed theoretical
study of the static spin susceptibility of 2D hole systems, which reveals unexpectedly rich
magnetic properties of p-type quantum wells. The paramagnetic response is
characterized by a strongly anisotropic and density-dependent effective
\textit{g}-factor. In a 2D DMS system, valence-band mixing drives magnetic transitions
that could enable new magneto-electronic device
functionalities~\cite{wolf:sci:01,flat:nphys:06} based on electric-field manipulation
of the magnetization in low-dimensional systems.

\textit{Background \& Aim} --
The magnetic properties of a many-particle system are most comprehensively characterized
by the spin-susceptibility tensor~\cite{yosidabook}. In a homogeneous electron system
with spin-rotational invariance, it has a universal isotropic form that depends only on the
dimensionality of the system~\cite{vignalebook}. This case applies to the 2D electron
systems realized by confining carriers from the conduction band in semiconductor
heterostructures as long as inversion symmetry is not broken by the crystal lattice or due
to structuring of the sample~\cite{ima:prb:04,Pletyukhov2007,mross:prb:09,ches:prb:10}.

\begin{figure}[b]
\includegraphics[width=0.8\columnwidth]{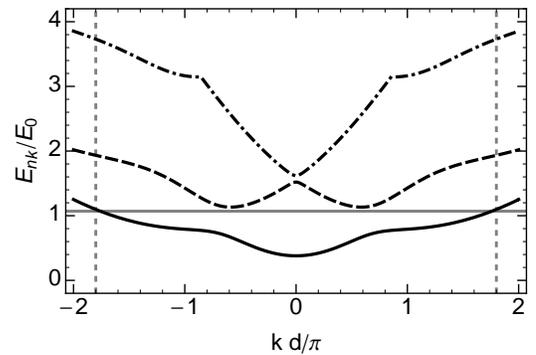}
\caption{\label{fig:dispersion}
Lowest three (each of them doubly degenerate) subbands of a two-dimensional hole system
realized by a symmetric hard-wall quantum-well confinement of width $d$. Dispersions are
calculated based on the four-band Luttinger-model description of bulk valence-band states in
axial approximation, using band-structure parameters applicable to GaAs confined in [001]
direction, and $E_0=-\pi^2\hbar^2 \gamma_1/(2 m_0 d^2)$. Gray lines indicate the range of
energies and wave vectors for which only the lowest subband is occupied. This is the regime
we focus on in this work.}                                                 
\end{figure}

Here we consider the properties of 2D \emph{hole\/} systems whose charge carriers have a
spin-3/2 degree of freedom that is strongly coupled to their orbital motion even when
inversion symmetry is intact~\cite{rolandbook}.  We adopt the Luttinger model~\cite{lutt:pr:56}
in axial approximation~\cite{rolandbook,suz:prb:74,treb:prb:79}, which  provides a useful
description of the upper-most valence band of typical semiconductors in situations where its
couplings to the conduction band and split-off valence band are irrelevant. 
Subband k-dot-p theory~\cite{broido:prb:85a,broido:prb:85b} is employed
to obtain the lowest quasi-2D hole subbands for a symmetric hard-wall confinement
characterized by its spatial width $d$. See Fig.~\ref{fig:dispersion}. We consider the case
where only the lowest 2D subband is occupied and calculate the spin susceptibility. Based
on this result, we discuss the paramagnetic response of 2D holes and identify various
magnetic phases that emerge in DMS quantum wells.

\textit{Spin susceptibility of a quasi-2D system\/} --
The spin susceptibility of 2D charge carriers is given by \cite{vignalebook}
\begin{eqnarray}
&& \chi_{ij}(\vek{R}, z; \vek{R}', z')= \nonumber \\ && \hspace{0.5cm}
-\frac{i}{\hbar} \int_0^\infty d t \,\, \ee^{-\eta t} \,\, \langle[S_i(\vek{R},z; t)\, ,
\, S_j(\vek{R}',z';0)] \rangle \, .
\label{eq:Spinsus}
\end{eqnarray}
Here $\vek{R}$ and $z$ are components of the position vector in the 2D ($xy$) plane
and in the perpendicular (growth) direction, respectively. The spin density operator
(in units of $\hbar$) is defined in terms of field
operators $\Psi$, $\Psi^\dagger$ and the Cartesian components $\hat J_j$ of the
charge carriers' intrinsic angular momentum as
$ S_j(\vek{R}, z )=\Psi^\dagger(\vek{R}, z)~\hat{J}_j~ \Psi(\vek{R}, z)$.
The field operators can be expressed in terms of operators associated with
general eigenstates [labelled by band index $n$ and in-plane wave vector $\vek{k}=
(k_x,k_y)$] of the non-interacting
Hamiltonian as
$\Psi(\vek{R}, z)=\sum_{n} \int \frac{d^2 k}{(2\pi)^2} \,\,\, \ee^{i \vek{k} \cdot \vek{R}}\,\,
\xi_{n \vek{k}}(z) \,\, c_{n\vek{k}}$.
The (normalized) spinors $\xi_{n\vek{k}}(z)$ and eigenvalues $E_{n\vek{k}}$ are
obtained by solving the multi-band Schr\"odinger equation for the confinement in growth
direction of the 2D heterostructure. We can then express the spin susceptibility as
\begin{subequations}\label{eq:SpinSTen}
\begin{equation}
\chi_{ij}(\vek{R}, z; \vek{R}', z') = \int \frac{d^2 q}{(2\pi)^2} \,\, \ee^{i \vek{q} \cdot (\vek{R} -
\vek{R}')} \,\, \chi_{ij}(\vek{q}; z, z') 
\end{equation}
in terms of the 2D Fourier-transformed susceptibility
\begin{eqnarray}
&& \chi_{ij}(\vek{q}; z, z') = \sum_{n,l} \int\frac{d^2 k}{(2\pi)^2} \,\, \mathscr{W}_{ij}^{nl}(\vek{k},
\vek{q}; z,z') \nonumber \\ &&\hspace{3.5cm} \times  \frac{n_{\text F}(E_{l\vek{k}
+\vek{q}})-n_{\text F}(E_{n \vek{k}})}{E_{l\vek{k}+\vek{q}} -E_{n\vek{k}}-i\hbar\eta}~,
\end{eqnarray}
where $n_{\text{F}}$ denotes the Fermi function, and
\begin{eqnarray}\label{eq:SpinOverlap}
&&\mathscr{W}_{ij}^{nl}(\vek{k},\vek{q}; z,z')= \Big[ \xi_{n\vek{k}}(z)\Big]^\dagger \cdot \Big[
\hat{J}_i \, \xi_{l\vek{k}+\vek{q}}(z)\Big] \nonumber \\ && \hspace{3cm} \times \Big[
\xi_{l\vek{k}+\vek{q}}(z')\Big]^\dagger \cdot \Big[ \hat{J}_j \, \xi_{n\vek{k}}(z')\Big] \,\, .
\end{eqnarray}
\end{subequations}
Axial symmetry of the 2D system implies $E_{n\vek{k}} \equiv E_{n k}$ (where
$k\equiv |\vek{k}|$ and $\phi_{\vek{k}}$ are the polar coordinates of $\vek{k}$) and
permits the ansatz~\cite{zhu:prb:87,zhu:prb:88}
\begin{equation}\label{eq:SepAnsatz}
\xi_{n\vek{k}}(z)=\ee^{-i\hat J_z \phi_{\vek{k}}} \, \, \bar\xi_{n k}(z) \quad ,
\end{equation}
simplifying calculation of the matrix elements $\mathscr{W}_{ij}^{nl}$.
In the following, we consider the growth-direction-averaged spin susceptibility
$\overline{\chi}_{ij} (\vek{q}) =  \int d z \int dz' \,\,\,\, \chi_{ij}(\vek{q}; z, z')$ calculated at
zero temperature.

\textit{Luttinger-model description of quasi-2D holes\/} -- Using the $4\times 4$
Luttinger Hamiltonian in axial approximation for the bulk valence band, the
bound states of 2D holes confined by a potential $V(z)$ are given in terms of spinor wave
functions $\bar\xi_{n k}(z)$ that satisfy the Schr\"odinger equation $\left[ \mathscr{H}_0 +
\mathscr{H}_1 + \mathscr{H}_2 \right] \bar\xi_{n k} = E_{n k}\, \bar\xi_{n k}$, where
\begin{subequations}
\begin{eqnarray}\label{eq:zoneCentHam}
\mathscr{H}_0 &=& \frac{\hbar^2}{2 m_0} \left[ \gamma_1\, \openone - 2 \tilde\gamma_1
\left( \hat J_z^2 -\frac{5}{4}\,\openone \right) \right] \frac{d^2}{dz^2} + V(z) \,\, , \\
\mathscr{H}_1 &=& \frac{\hbar^2 k}{m_0} \, \sqrt{2} \,\tilde \gamma_2\, (-i) \left( \{\hat J_z ,
\hat J_- \}+ \{\hat J_z , \hat J_+ \} \right) \frac{d}{d z} \,\, , \\
\mathscr{H}_2 &=& -\frac{\hbar^2 k^2}{2 m_0}\left[ \gamma_1\, \openone + \tilde\gamma_1
\left( \hat J_z^2 -\frac{5}{4}\,\openone \right) - \tilde\gamma_3 \left( \hat J_+^2 + \hat J_-^2
\right)\right] . \nonumber \\
\end{eqnarray}
\end{subequations}
Here $m_0$ is the electron mass in vacuum, hole energies are counted as negative from
the bulk valence-band edge, and we used the abbreviations $\hat J_\pm = (\hat J_x \pm i
\hat J_y) / \sqrt{2}$, $\{ A, B\} = (A B + B A)/2$. The constants $\gamma_1$ and $\tilde
\gamma_j$  are materials-related band-structure parameters~\cite{vurg:jap:01} and depend 
on the quantum-well growth direction.

Straightforward application of the subband k-dot-p method ~\cite{broido:prb:85a,broido:prb:85b}
yields the energy dispersions $E_{nk}$ with associated eigenspinors $\bar\xi_{nk}$. 
At $k=0$, the eigenspinors are also eigenstates of $\hat J_z$ with eigenvalues $\pm 3/2$
(heavy holes) or $\pm 1/2$ (light holes), which are split in energy. This phenomenon is often
referred to as \emph{HH-LH splitting\/}~\cite{suz:prb:74,rolandbook}. At finite $k$, the spinors
$\bar\xi_{nk}$are not eigenstates of $\hat J_z$ anymore. This phenomenon of
\emph{HH-LH mixing\/} arises because a spin-3/2 degree of freedom has a much richer
structure than the more familiar spin-1/2 case~\cite{roland:prb:04}.
Figure~\ref{fig:dispersion} shows the 2D-subband dispersions obtained for a symmetric
hard-wall confinement of width $d$, using band-structure parameters for GaAs confined
in [001] direction. The numerically obtained $E_{nk}$ and $\bar\xi_{nk}$ serve as input for
the calculation of the 2D-hole spin susceptibility according to Eqs.~(\ref{eq:SpinSTen}) with 
Eq.~(\ref{eq:SepAnsatz}).

\begin{figure}[t]
\includegraphics[width=0.9\columnwidth]{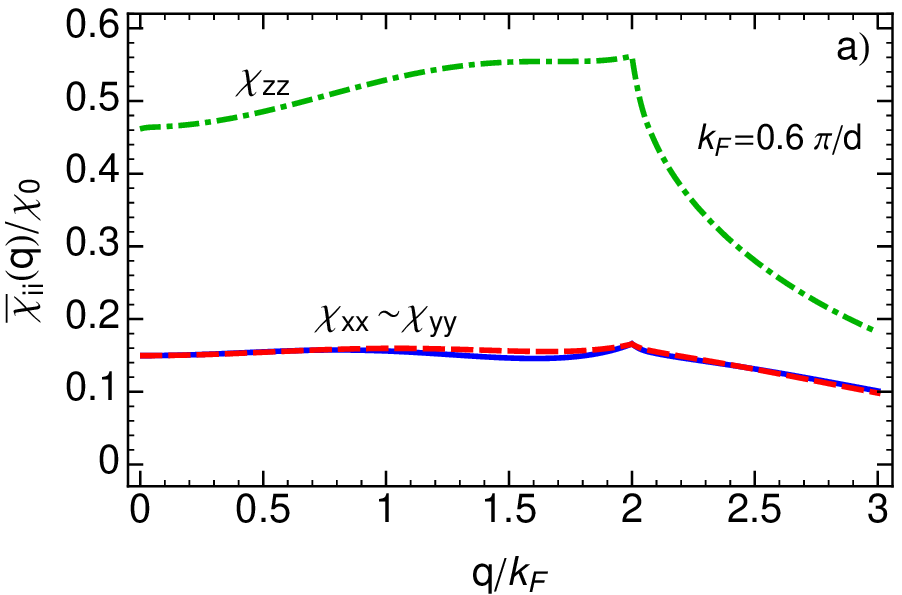}\\
\includegraphics[width=0.9\columnwidth]{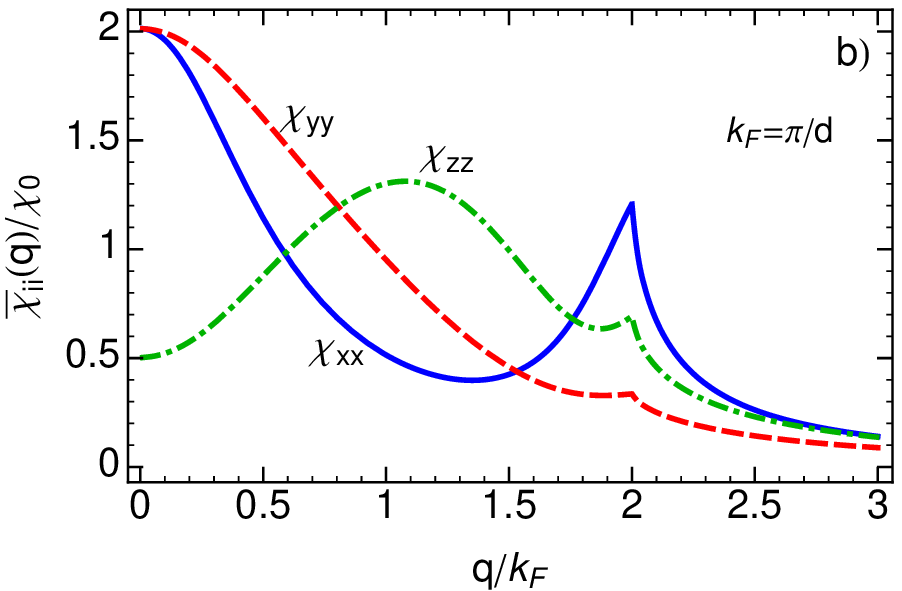}\\
\includegraphics[width=0.9\columnwidth]{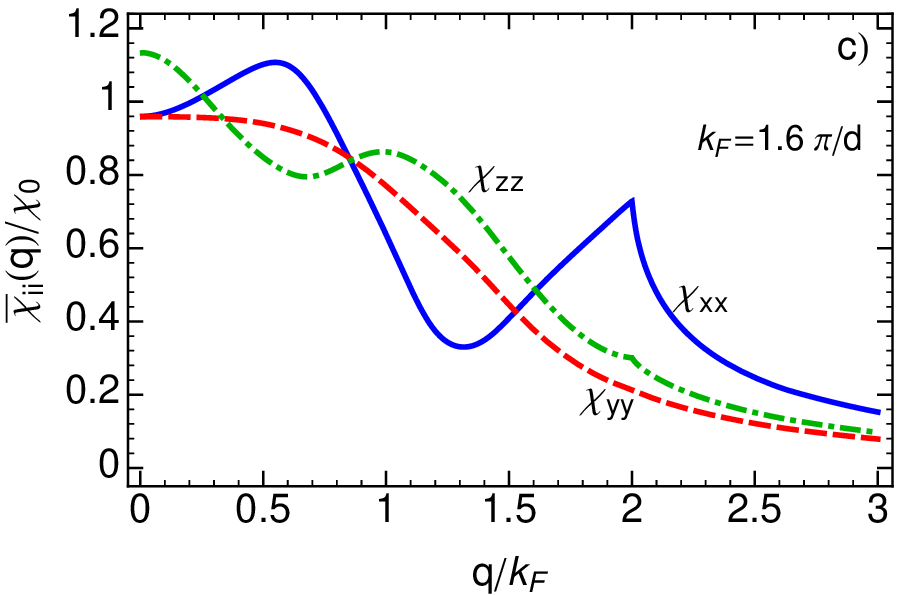}
\caption{\label{fig:chiq}
$\vek{q}$-dependent spin susceptibility for  $\phi_\vek{q}=0$ and values of the 2D-hole
Fermi wave vector $k_{\text{F}}$ as indicated in the individual panels. Calculations are
based on the band structure shown in Fig.~\ref{fig:dispersion}.}
\end{figure}

\textit{Results for the 2D-hole spin susceptibility\/} -- 
Using polar coordinates $(q,\phi_\vek{q})$ for the 2D wave vector $\vek{q}$ and introducing
the scale $\chi_0 = 2 m_0/(\gamma_1 \hbar^2)$, we find  that the tensor elements of
$\overline{\chi}_{ij}({\bf q})$ have the generic form
\begin{subequations}\label{eq:GenStruc}
\begin{eqnarray}
\label{eq:GenStrucxx}
\overline{\chi}_{xx}(\vek{q}) &=& \chi_0\left[ F_\parallel(q)+G(q)
\, \cos\left(2\phi_\vek{q}\right) \right] ,\\
\overline{\chi}_{yy}(\vek{q}) &=& \chi_0\left[F_\parallel(q) - G(q)
\, \cos\left(2\phi_\vek{q}\right) \right] ,\\
\overline{\chi}_{xy}(\vek{q})&=& \chi_0\, G(q) \, \sin\left( 2\phi_\vek{q}\right)
\,\, ,\\ \label{eq:GenStruczz}
\overline{\chi}_{zz}(\vek{q})&=& \chi_0\, F_\perp(q) \,\,.
\end{eqnarray}
\end{subequations}
The functions $F_{\parallel,\perp}(x)$  and $G(x)$ depend on materials and morphological
parameters of the 2D hole system, especially the hole sheet density $n_{\text{2D}}\equiv
k_{\text{F}}^2/(2\pi)$, but $G(0)=0$ generally. Figure~\ref{fig:chiq} shows typical results
obtained at low, intermediate and high densities where still only states in the lowest
quasi-2D subband are occupied \footnote{See the Supplementary Material for the
corresponding real-space structure of the 2D-hole spin susceptibility, which is relevant
for the proposed use of RKKY interaction to couple quantum-dot confined
spins~\cite{ima:prb:04,mross:prb:09,ches:prb:10}.}.

The density dependence of the 2D-hole spin susceptibility follows a generic trend. In the
limit of low density [Fig.~\ref{fig:chiq}(a)], $\overline{\chi}_{zz}({\bf q})\gg \overline{\chi}_{xx}
({\bf q})\sim \overline{\chi}_{yy}({\bf q})$, and the line shape of $\overline{\chi}_{zz}({\bf q})$ 
is similar to the universal 2D-electron (Lindhard) result~\cite{vignalebook}. The strong
easy-axis response is expected~\cite{haury:prl:97,dietl:prb:97,wurst:nphys:10} as a result of
HH-LH splitting, which favors a spin-3/2 quantization axis perpendicular to the 2D
plane~\cite{rolandbook}. Interestingly, the behavior of the spin susceptibility changes as
density is increased. For intermediate values of hole density, results like that shown in
Fig.~\ref{fig:chiq}(b) are obtained, exhibiting \emph{easy-plane\/} anisotropy in the
long-wave-length limit and a nontrivial structure developing at wave vectors comparable to
$k_{\text{F}}$. The significant deviation from both the universal-2D-electron behavior and
also the easy-axis response expected from HH-LH \emph{splitting\/} arises because, as the
2D-hole density is increased, higher-$k$ states get occupied that are more strongly
influenced by HH-LH \emph{mixing\/}. At the highest values of density where still only the
lowest quasi-2D hole subband is populated, easy-axis anisotropy is restored in the
long-wave-length limit but the response at finite $q$ becomes as important in strength as
that for $q\to 0$.

\textit{Effective g-factors for quasi-2D holes} --
Motivated by recent experimental~\cite{chi11,bou06,win05a} and theoretical~\cite{win05}
interest in the paramagnetic response of 2D holes, we apply our results for the spin
susceptibility of a noninteracting 2D hole system to define an effective \textit{g} factor. 

\begin{figure}[b]
\includegraphics[width=0.9\columnwidth]{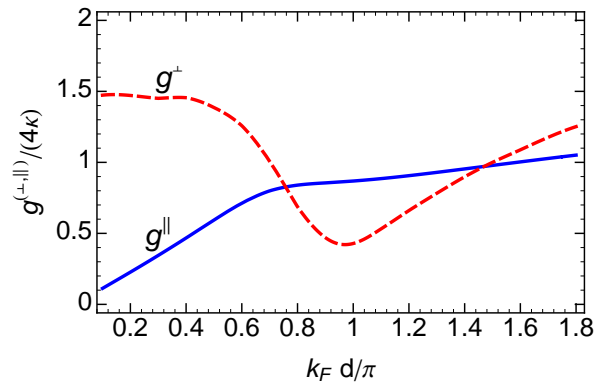}
\caption{\label{fig:gj}
Effective $g$-factor  as a function of the 2D-hole Fermi wave vector $k_{\text F}$ for a
perpendicular (dashed curve) and in-plane (solid curve) magnetic field.}
\end{figure}

A magnetic field parallel to the $j$ axis couples to the holes' spin via the Zeeman term
$\mathscr{H}_{\text{Z}} = 2\kappa\, \mu_{\text{B}}\, {B_j}\,\hat{J}_j $, where $\mu_{\text{B}}$
is the Bohr magneton and $\kappa$ the bulk-hole \textit{g} factor~\cite{rolandbook}. In the
low-field limit, the paramagnetic susceptibility is given by $\chi_{\text{P},j} = \left( 2 \kappa
\,\mu_{\text{B}} \right)^2 \, \overline{\chi}_{jj}(\vek{q}=0)$ in terms of our calculated spin
susceptibility~\footnote{The paramagnetic-susceptibility tensor is diagonal for the case
under consideration, reflecting the structure of the 2D-hole spin susceptibility tensor
Eqs.~(\ref{eq:GenStruc}) at $\vek{q}=0$.}. Comparison of this relation with the expression
for the Pauli susceptibility of conduction electrons from a parabolic band~\cite{vignalebook}
suggests defining the effective $g$-factor for a many-particle state via $\chi_{\text{P},j}
\equiv \left( \frac{g_j \mu_{\text{B}}}{2} \right)^2\overline{\chi}_{\text{L}}(\vek{q}=0)$, which
explicitly yields
\begin{equation}\label{eq:effHole}
g_j = 4\kappa \, \sqrt{\frac{\overline{\chi}_{jj}(\vek{q}=0)}{\overline{\chi}_{\text{L}}(\vek{q}=0)}}
\quad .
\end{equation} 
Here $\overline{\chi}_{\text{L}}$ is the static 2D-hole Lindhard function; its $\vek{q}=0$
limit equals the density of states at the Fermi level. We find a density-dependent and
anisotropic $g$-factor, reflecting the interplay between HH-LH splitting and mixing in
confined valence-band states. Figure~\ref{fig:gj} shows the density dependence of the
transverse $(\perp)$ and in-plane $(\parallel)$ $g$-factors for a 2D hole system. In the
low-density limit, behavior expected from HH-LH splitting is found, whereas the
paramagnetic response at intermediate and high density is substantially affected by
HH-LH mixing.

\textit{2D-hole-mediated magnetism} --
Magnetism is introduced into intrinsically nonmagnetic semiconducting materials via doping
with magnetic ions~\cite{ohno:sci:98,jung:rmp:06,dietl:nmat:10} such as Mn, Co, Fe or Gd.
One way to generate an exchange interaction between any two localized magnetic moments
embedded in a conductor is provided by the Ruderman-Kittel-Kasuya-Yosida (RKKY)
mechanism~\cite{yosidabook}, which gives rise to the effective two-impurity spin Hamiltonian
\begin{equation}\label{eq:RKKYHam}
\mathscr{H}_{\alpha\beta}=-G^2 \sum_{i,j} I^{(\alpha)}_{i} I^{(\beta)}_{j} \chi_{ij}
({\vek{R}_\alpha,z_\alpha; \vek{R}_\beta,z_\beta}) \quad .
\end{equation}
Here $I_{i}^{(\alpha)}$ denotes the $i$th Cartesian component of an impurity spin located at
position $(\vek{R}_\alpha,z_\alpha)$, and $G$ is the exchange constant for the contact interaction
between the spin density of delocalized charge carriers with the impurity spins. For a random but
on average homogeneous distribution of magnetic ions, standard mean-field
theory~\cite{yosidabook} applied to the spin model described by the Hamiltonian of
Eq.~(\ref{eq:RKKYHam}) yields the Curie temperatures 
\begin{subequations}
\begin{equation}\label{eq:CurieT}
T_{\text{C}_j}^{\text{(MF)}} =  8\pi~T_0~ \frac{\overline\chi_{jj}(\vek{q}=0)}{\chi_0}
\end{equation}
for ferromagnetic order with magnetization direction parallel to the $j$ axis. The temperature
scale
\begin{equation}\label{eq:temp0}
T_0 = \frac{I(I+1)}{12}\, \frac{G^2}{k_{\text{B}}}\, \frac{n_I}{d}\, \frac{m_0}{\pi\hbar^2 
\gamma_1}
\end{equation}
\end{subequations}
depends on the impurity-spin magnitude $I$ and the average 3D density $n_I$ of magnetic
impurities, and its functional form is that obtained for 2D charge carriers in a parabolic
band~\cite{dietl:prb:97,lee:prb:00} with effective mass $m_0/\gamma_1$.

\begin{figure}[b]
\includegraphics[width=0.9\columnwidth]{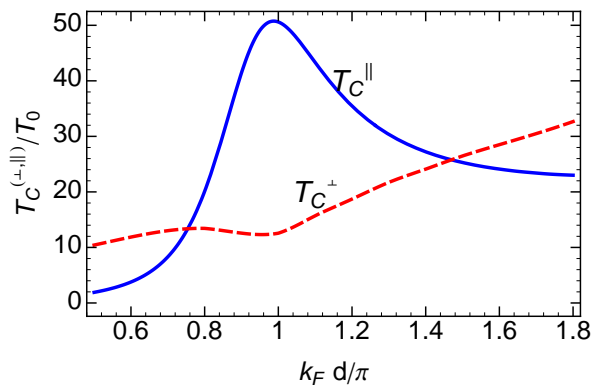}
\caption{\label{fig:curieT}
Mean-field Curie temperatures for 2D-hole-mediated easy-axis ($T_{\text{C}}^{\perp}$)
and easy-plane ($T_{\text{C}}^{\parallel}$) magnetism as a function of the 2D-hole Fermi
wave vector $k_{\text{F}}$, obtained for a hard-wall confinement with width $d$ and
band-structure parameters applicable to GaAs.}
\end{figure}

Figure~\ref{fig:curieT} shows the mean-field Curie temperatures for perpendicular-to-plane
($\perp$) and in-plane ($\parallel$) magnetization directions calculated with the same input
parameters used for obtaining the subbands given in Fig.~\ref{fig:dispersion}. The density
dependence of $\overline\chi_{xx,zz}(0)$ is directly reflected in that of
$T_{\text{C}}^{\parallel,\perp}$. At the mean-field level, the ordered state associated with the
maximum transition temperature will be established.
Our results suggest that 
the type of magnetic ordering can be modified by changing the 2D-hole density, e.g., by
adjusting the gate voltage in accumulation-layer devices~\cite{che12}. Easy-axis
magnetism prevails at low and high densities, whereas an unexpected easy-plane
magnetic order emerges at intermediate values of the density. Also for high densities, a
local maximum appears in $\overline{\chi}_{xx}({\bf q})$ at $q\approx 0.6 k_{\text{F}}$, which
almost reaches the value of $\overline{\chi}_{zz}(q=0)$. See Fig.~\ref{fig:chiq}(c). Even after 
averaging over the polar angle, we find that for $k_{\text{F}}\sim 1.5~\pi/d$ the in-plane
susceptibility can have a maximum at $q\neq 0$ which is as large as $\overline{\chi}_{zz}
({q=0})$. Thus it may be possible that the high-density easy-axis ferromagnetic state must
coexist (or compete) with helical magnetism~\cite{yosidabook}.

\textit{Finite-temperature effects} -- 
Thermal excitation of spin waves (magnons) suppresses the magnitude of the magnetization
below its mean-field value $M_0$. 
This effect is captured by the relation~\cite{sim:prb:08,zak:prb:12}
\begin{equation}
\frac{M(T)}{M_0} = 1 -\frac{1}{4\pi^2 n_I d} \int d^2 q \,\,\, n_{\vek{q}}\big(T\big)\quad ,
\label{eq:magnetisation}
\end{equation}
where $n_{\vek{q}}(T)$ is the occupation-number distribution function of magnon
modes at temperature $T$. A spin-wave-related critical temperature is defined by the
condition $M(T_{\text{C}}^{\text{(SW)}})=0$ because, for $T>T_{\text{C}}^{\text{(SW)}}$,
too many magnon excitations will have been excited to sustain a finite magnetization.
In equilibrium, $n_{\vek{q}}(T)$ is given by the Bose-Einstein distribution function
$n_{\text{B}}(\varepsilon_\vek{q})=1/(\ee^{\varepsilon_\vek{q}/[k_{\text{B}} T]} - 1)$, which
depends on the spin-wave energy dispersion $\varepsilon_\vek{q}$. The latter's expression
in terms of the charge carriers' $\vek{q}$-dependent spin susceptibility depends on the
type of magnetic order (Heisenberg, Ising, or helical)~\cite{sim:prb:08}, but the
parameterisation
\begin{equation}
\varepsilon_\vek{q} = I\, G^2 \,\frac{n_I}{d}\,\chi_0 \left[ \bar \varepsilon_0 + \bar c_\nu\,
\left(\frac{q}{k_{\text{F}}}\right)^\nu\right]
\label{eq:MagnonDisAnsatz}
\end{equation}
typically holds for the relevant energy range. The dimensionless quantities
$\bar\varepsilon_0$, $\nu$, $\bar c_\nu$ can be determined from the functional form
of the $\vek{q}$-dependent spin susceptibility. Stability of the magnetic  order requires
both coefficients $\bar\varepsilon_0$ and $\bar c_\nu$ to be positive. Specializing to
our situation of 2D-hole-mediated magnetism, we see from Fig.~\ref{fig:chiq}(a) that
the easy-axis magnetism expected at low hole-sheet densities is destabilized by
magnons because $\varepsilon_\vek{q}<0$. Considering the easy-plane magnetism
at intermediate densities, Fig.~\ref{fig:chiq}(b) reveals that the associated magnon
dispersion is characterized by $\nu=2$ and $\bar \varepsilon_0=0$, which again
implies destabilization of this magnetic order due to spin-wave excitations. For the
easy-axis (Ising) magnet expected at high densities [cf. Fig.~\ref{fig:chiq}(c)], we
find $\nu=2$ and $\bar\varepsilon_0>0$. In this case a finite spin-wave-related
critical temperature is obtained. Further studies need to explore the effect of
Coulomb interactions, which can stabilize ferromagnetic order mediated by 2D
carriers~\cite{zak:prb:12}.

\textit{Conclusions\/} --
The spin susceptibility of 2D holes is strongly density dependent. In the low-density
limit, the easy-axis response due to HH-LH \textit{splitting\/} is exhibited. With
increasing density, HH-LH \textit{mixing\/} changes the spin-related response of
confined holes even more drastically than the density
response~\cite{cheng:prb:01,lopez2001,kern10}. An effective \textit{g} factor for
2D holes is proposed. We clarify the impact of band-structure effects in 2D DMS
systems that had previously been only considered for the 3D
case~\cite{fiet:prb:05,timm:prb:05,kyry:prb:11} or outside the RKKY
limit~\cite{lee:sst:02}. The switching behavior of the magnetization found here
should be observable in p-type quantum wells where the 2D-hole density is
independently adjustable~\cite{che12} and the magnetic doping is sufficiently low
to ensure a sizable mean free path.

\textit{Acknowledgments} --
The authors benefited from useful discussions with A.~H.~MacDonald, J.~Splettst\"o{\ss}er,
and R. Winkler.


%

\clearpage
\onecolumngrid
\renewcommand{\thefigure}{S\arabic{figure}}
\setcounter{figure}{0}

\section*{Supplementary Material for ``Carrier-Density-Controlled Anisotropic Spin Susceptibility of Two-Dimensional Hole Systems"}

\begin{figure}[h]
\includegraphics[width=0.45\columnwidth]{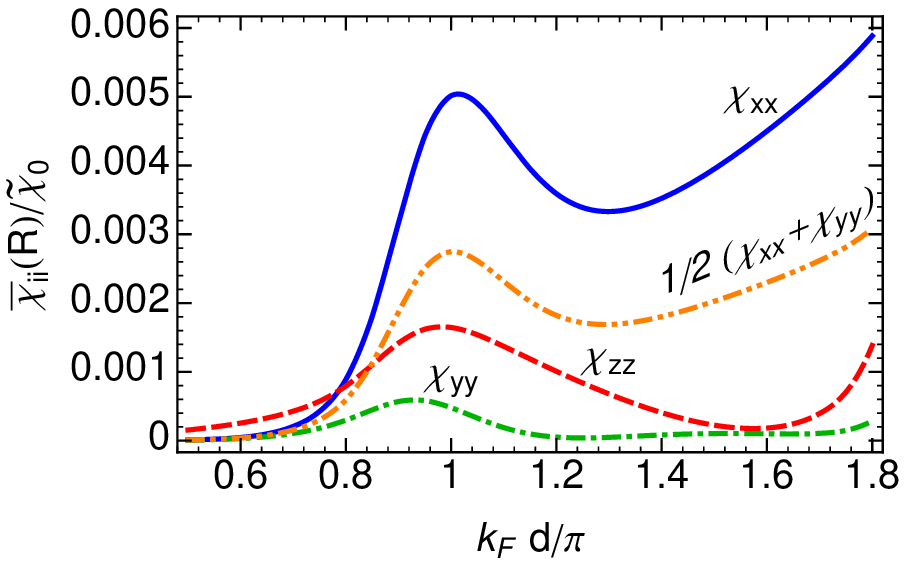}
\caption{\label{fig:chiComps}
Real-space spin-susceptibility $\overline{\chi}_{ij}(\vek{R})$ of a 2D hole system, plotted
in units of $\tilde{\chi}_0=2\pi^2 m_0/(\gamma_1 \hbar^2 d^2)$ as a function of $k_{\text{F}}
d$ for fixed $\phi_\vek{R}=0$ and $k_{\text{F}} R=10$. [Here $(R,\phi_{\vek{R}})$ denote the
polar coordinates of the 2D real-space vector $\vek{R}$.]}
\end{figure}

Fourier transforming $\overline{\chi}_{ij}({\bf q})$ yields the spin susceptibility in real space
$\overline{\chi}_{ij}({\bf R})$, which 
is relevant, e.g., for the proposed use of RKKY interactions to couple
quantum-dot-confined spins as part of quantum-information-processing
protocols~\cite{ima:prb:04,mross:prb:09,ches:prb:10}. Figure~\ref{fig:chiComps} shows
the density dependence of $\overline{\chi}_{ij}({\bf R})$ for a 2D hole system. 
In the low-density
limit, we find $\overline{\chi}_{zz}\gg \overline{\chi}_{xx}\sim\overline{\chi}_{yy}\sim 0$,
reflecting the strong easy-axis anisotropy arising due to HH-LH
splitting~\cite{haury:prl:97,dietl:prb:97}. As the density of 2D holes in the lowest subband
increases, in-plane components are observed to become relevant. For $k_{\text F} \sim
1.6 \pi/d$ we find $\overline{\chi}_{xx}$ to be the dominant component of the
spin susceptibility tensor while $\overline{\chi}_{zz}\sim\overline{\chi}_{yy}\sim 0$.
Switching of the effective
Hamiltonian that describes the coupling of two localized spins at a fixed distance is
therefore possible by simply adjusting the density of the 2D hole system in which they
are embedded.

\end{document}